\documentclass[prl,showpacs,preprint]{revtex4-1}

\usepackage{amsmath}
\usepackage{amssymb}
\usepackage{times}
\usepackage{wasysym}
\usepackage{graphicx}
\usepackage{psfrag}
\usepackage{color}
\usepackage{cancel}

\begin{document}

\title{Critical Line of the O($N$) Loop Model on
  the Square Lattice}

\author{Ant\^onio M\'arcio P. Silva, Adriaan M. J.~Schakel, and Giovani L. Vasconcelos}

\affiliation{
Laborat\'orio de F\'isica Te\'orica e Computacional,
Departamento de F\'{i}sica,
Universidade Federal de Pernambuco, 
50670-901, Recife-PE, Brazil}

\begin{abstract}
 An efficient algorithm is presented to simulate the $O(N)$ loop model
 on the square lattice for arbitrary values of $N>0$.  The scheme
 combines the worm  algorithm with a new data structure to
 resolve both the problem of loop crossings and the necessity of counting the number of loops at each Monte Carlo update. With the use of
 this scheme, the line of critical points (and other
 properties) of the O($N$) model on the square lattice for $0<N\le 2$
 have been determined.
\end{abstract}

\maketitle

The self-avoiding walk together with the Ising and XY models formulated
on the square lattice are among the most studied models in statistical
physics, with many applications in diverse fields.  These spin systems
belong to the class of $N$-vector models with $N=0$, $1$, and $2$,
respectively. Their analytic continuation to arbitrary $N$, with $-2
\leq N \leq 2$, was, until recently, almost exclusively studied on the
honeycomb lattice instead.  These models undergo a
continuous phase transition between a disordered high-temperature phase
and an ordered low-temperature phase.  Studies for non-integer $N$ are
facilitated by the high-temperature (HT) representation of the spin
system where contributions to the partition function are represented by
even graphs along the links of the underlying lattice.  On a
honeycomb lattice, even graphs cannot cross and thus automatically form
mutually and self-avoiding polygons.  This fact greatly simplifies the
analysis and explains the ``exorbitant amount of work'' done on the
honeycomb lattice \cite{CPS}.  In most of these studies it was tacitly
assumed that the long-distance behavior found on the honeycomb lattice
is universal and insensitive to the lattice structure. This turned out
to be true for the long-distance behavior controlled by the O($N$)
critical point with $-2 \leq N \leq 2$ found in Ref.~\cite{Nienhuis}.
But in the low-temperature phase, which also displays long-range
spin-spin correlations with powerlike decay, the situation changes
dramatically.  By working on a square lattice, Jacobsen \emph{et. al.}
\cite{JRS} argued that allowing for crossings gives rise to a completely
different long-distance behavior from that found on a honeycomb lattice
\cite{Nienhuis}.  Although several numerical studies of O($N$) models
for non-integer $N$ on the square lattice recently appeared
in the literature, see, for example, \cite{Deng2007,Guo2011}, a proper
account of crossings has not yet been accomplished.

\begin{figure} 
\begin{center}
\includegraphics[width=.40\textwidth]{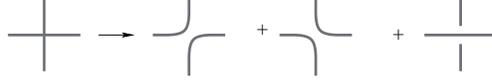} 
\end{center}
\caption{Resolving a crossing (left)  on a
  square lattice.  The intersections on the right come with unambiguous
  routings.}
\label{fig:crossing} 
\end{figure}

This task is hindered by two main obstacles. The first one is that each even HT graph
must be decomposed into an unambiguous set of polygons or {\it loops},
with each loop carrying a degeneracy factor $N$ that arises because any
of the $N$ spin components or colors can be routed through the loop.
This problem was formally solved by Chayes \emph{et al.}~\cite{CPS} by
resolving each crossing of a HT graph into three routings, with unique
instructions as how to connect the four legs, see
Fig.~\ref{fig:crossing}. A given even graph configuration
then breaks into a set of interlocking colorless loops that can be
assigned unambiguous degeneracy factors.  To the best of our knowledge, this scheme has never been employed in numerical studies.

The second obstacle is that at each update step the number of loops
present on the entire lattice is required.  Since this quantity is
nonlocal, the use of a tracing algorithm would slow down simulations
in the critical region as well as in the low-temperature phase to
the extent of rendering them useless on larger lattices.

In this paper, we implement the scheme due to Chayes \emph{et
  al.}~\cite{CPS}  by adopting the worm algorithm
\cite{ProkofevSvistunov} and overcome the tracking problem
 by coding loops into a new data structure, called a
\emph{satellite list} \cite{satellite}.  As an application, we determine the line
of critical points of the O($N$) model on the square lattice for
arbitrary $0 < N < 2$.

Our starting point  is the partition function~\cite{DMNS}
\begin{equation} 
\label{ZS} 
Z_\square = \mathrm{Tr} \prod_{\langle i,j \rangle} (1 + K \mathbf{S}_i
\cdot \mathbf{S}_j) ,
\end{equation} 
obtained by truncating the series expansion of the conventional
Boltzmann factor, $\exp(K \mathbf{S}_i \cdot \mathbf{S}_j)$, at the
first order in the coupling constant $K$.  In Eq.~(\ref{ZS}), the
$N$-component spin variable $\mathbf{S}_i$ at each site $i$ is
normalized, $|\mathbf{S}_i|=1$, the trace Tr stands for the (normalized)
sum or integral over all possible spin configurations, and the product
is restricted to nearest neighbor pairs.  Notice that $K$ in
Eq.(\ref{ZS}) must satisfy the condition $|K| \leq 1$, if the weight of
a spin configuration is to be non-negative.  In the following, we take $K>0$.

In the HT representation of the model (\ref{ZS}), a factor $K
\mathbf{S}_i \cdot \mathbf{S}_j$ is represented by a bond drawn between
the nearest neighbor sites $i$ and $j$ which is given a weight $K$.  The
advantage of the truncated model (\ref{ZS}) compared to the conventional
one is that links cannot be multiply occupied.  Unlike crossings, this
simplification is believed to be irrelevant both at the critical point
and in the low-temperature phase.  (For recent numerical support of this
assumption, see \cite{Guo2011}.)  Each site carries a weight $Q(k_i)$
that depends on the number of bonds $k_i$ attached to that site.
Specifically, 
$ Q(0) = 1$, $Q(2)=1/N$,  and $Q(4) = 1/N(N+2)$ for sites
  connecting $0, 2$, or $4$ bonds \cite{CPS}.  The
partition function for $N>0$ can then be written as a sum over
configurations, $\mathcal{C}$, consisting of arbitrary many colorless
loops \cite{CPS}:
\begin{equation} 
\label{Zsquare} 
Z_\square = \sum_{\mathcal{C}} K^b \left(\frac{1}{N}\right)^{m_2}
\left(\frac{1}{N(N+2)}\right)^{m_4} N^\ell .
\end{equation} 
Here, $b$ denotes the total number of bonds, $K$ plays the role of a
bond fugacity, $m_2$ is the number of lattice sites connecting two
bonds, $m_4$ is the number of intersections (each with a unique routing
instruction), and $\ell$ is the number of loops in a given loop
configuration.  Finally, the degeneracy factor $N$ appears as a loop
fugacity in the HT representation.  Whereas in the spin representation
(\ref{ZS}) $K$ must satisfy the bound $K \leq 1$, no such restriction is
required in the HT representation~(\ref{Zsquare}) and $K$ can be
continued to the region $K > 1$.  

We simulate the model (\ref{Zsquare}) on a square lattice (with periodic
boundary conditions) for arbitrary $N>0$ by using the worm algorithm
\cite{ProkofevSvistunov}, a very efficient Monte Carlo
scheme that is not hampered by critical slowing down.  This scheme
directly generates HT graphs through the motion of an endpoint of an
open chain,  or \emph{worm}.  When the head
of the worm encounters a site that already connects two bonds, our
algorithm randomly chooses one of the three possible routings, so that
the resulting graphs are always unambiguous.  At each time step, the worm
attempts to either set or erase a bond, depending on whether the link
under consideration was previously empty or not.  As time proceeds, the
worm constantly changes its shape and length until its head eventually
returns to its tail to form a loop.  After this, a new worm is attempted
to be created by proposing to move both endpoints together to a randomly
chosen site and the process continues for as long as necessary to
accumulate sufficient statistics.  Notice that although the head shifts
smoothly through the lattice, taking steps of one lattice spacing at a
time, the worm can change its shape and length abruptly, either by
opening up and annexing an existing loop or by intersecting itself and
thereby shedding a loop.  The latter process is known in polymer physics
as {\it backbiting}. The Metropolis update probabilities for setting or
erasing a bond readily follow from detailed balance and will not be
discussed here for want of space, save for noting that the existence of
multiple routings at intersections complicates matters, for they can
lead to different degeneracy factors.

Whereas only loop configurations contribute to the
partition function $Z$,  configurations that include a
worm with endpoints $i_1$ and $i_2$ contribute to the unnormalized
two-point spin-spin correlation function $Z_{i_1 i_2}$.  This function
is related to the normalized correlation function through $\langle
\mathbf{S}_{i_1} \cdot {\mathbf S}_{i_2} \rangle = Z_{i_1 i_2}/Z$.
Because of the dot product appearing here, a worm, like a loop, carries
a degeneracy factor $N$.  For the truncated model, each lattice site on
a square lattice can connect only $0, 2$, or $4$ bonds, with the
exception of the sites housing an endpoint of the worm which connect $1$
or $3$ bonds.  In that case, the site $i$ hosting an additional endpoint
is assigned the weigth $Q(k_i+1)$ for $k_i =1,3$, where the extra
 term ``1'' in the argument is to indicate
explicitly the presence of an endpoint at that site.  
Since
$\mathbf{S}_i \cdot \mathbf{S}_i = 1$, a site $i$ housing both endpoints
at the same time acts as if they are absent, and $Z_{i i}=Z$.
Note that our convention here differs from that used in Ref.~\cite{ProkofevSvistunov}.
The critical exponents can be determined directly from the
  HT graphs through the use of observables known from
  percolation theory and the theory of self-avoiding random walks
  \cite{worms}.

\begin{figure}
\begin{center}
\includegraphics[width=.4\textwidth]{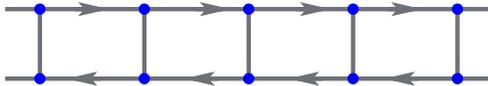} 
\end{center}
\caption{A satellite list: two singly-linked lists (the two sidepieces)
  running through the satellites (dots) of
  each node (crosspiece).}
\label{fig:ladder}
\end{figure}

We have overcome the problem of keeping track of the number of loops at
each Monte Carlo update by coding the worm (and the loops it generates)
into a satellite list \cite{satellite}.  Such a list, which is a
variation on the more familiar doubly-linked list, is a ladderlike data
structure, see Fig.~\ref{fig:ladder}.  A crosspiece in the ladder
represents a node where the relevant information (in our case, an
occupied link) is stored.  Each node contains two satellites,
represented by the endpoints of the corresponding crosspiece, and the
sidepieces represent singly-linked lists connecting the respective
satellites on the two sides of the ladder. In this way, nodes are linked
indirectly through their satellites in such a way that the next node in
the list depends on the current direction. The possibility to traverse
the list in one direction and, by sidestepping to the complement
satellite, also in the reverse direction is crucial for our purposes.
As with all linked lists, memory allocation is dynamic so that efficient
memory usage is guaranteed.

\begin{figure*}
\psfrag{x}[t][t][.7][0]{$K'$}
\psfrag{y}[t][t][.7][0]{$\Pi_L$}
\begin{center}
  \includegraphics[width=.29\textwidth]{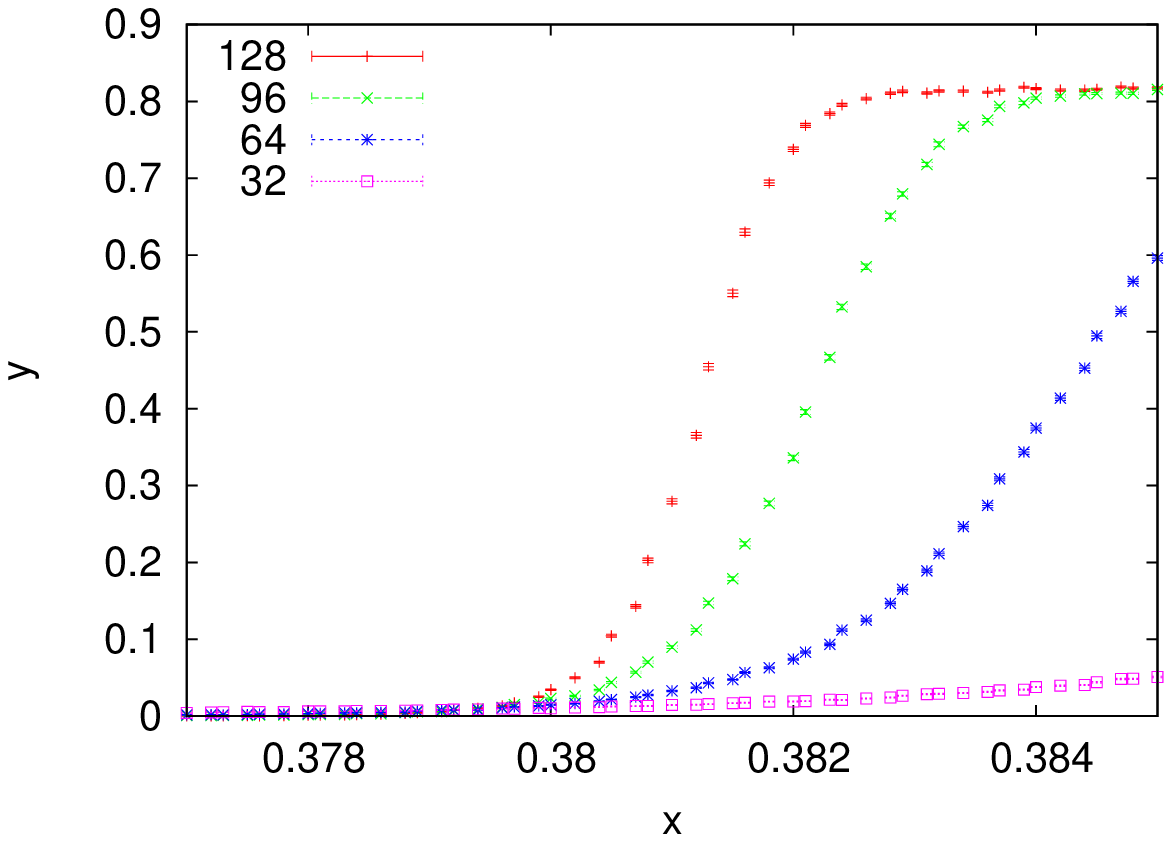} \hspace{.05\textwidth}
  \includegraphics[width=.29\textwidth]{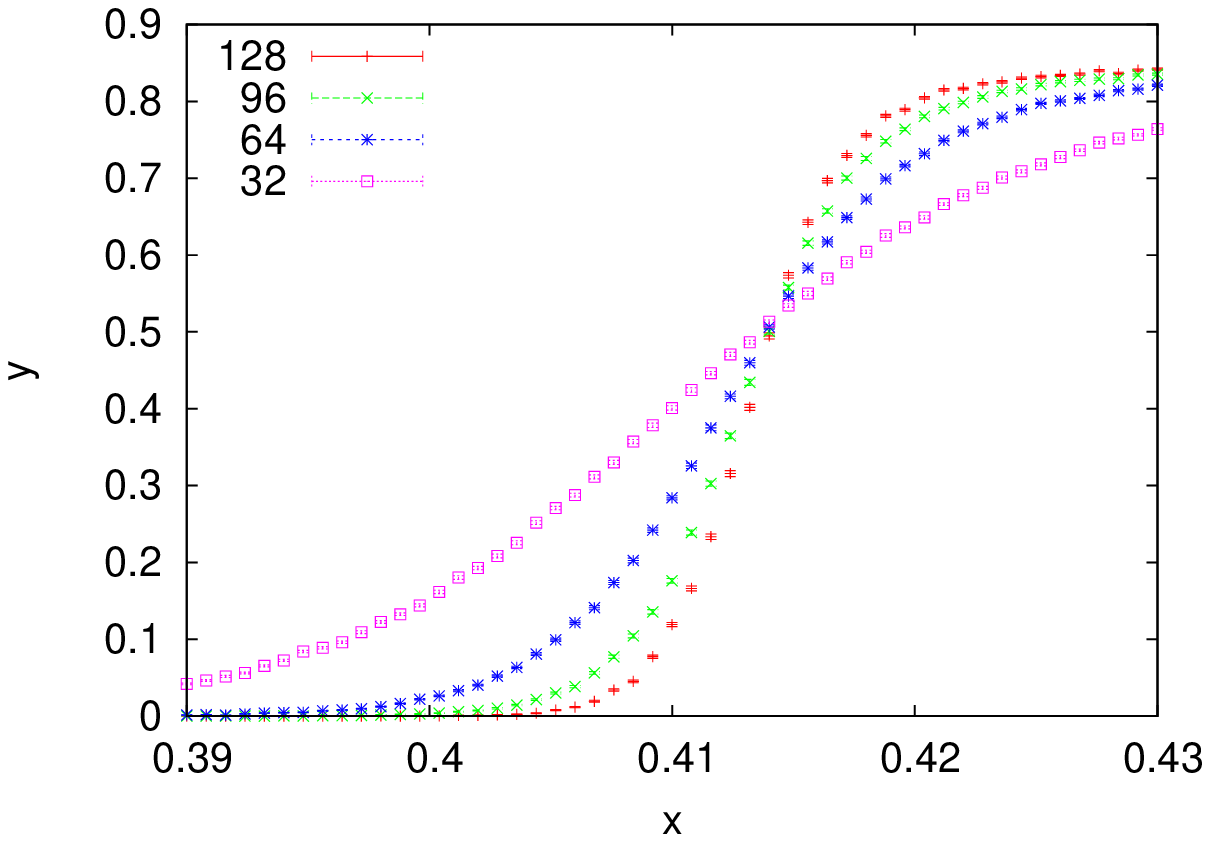}  \hspace{.05 \textwidth}
  \includegraphics[width=.29\textwidth]{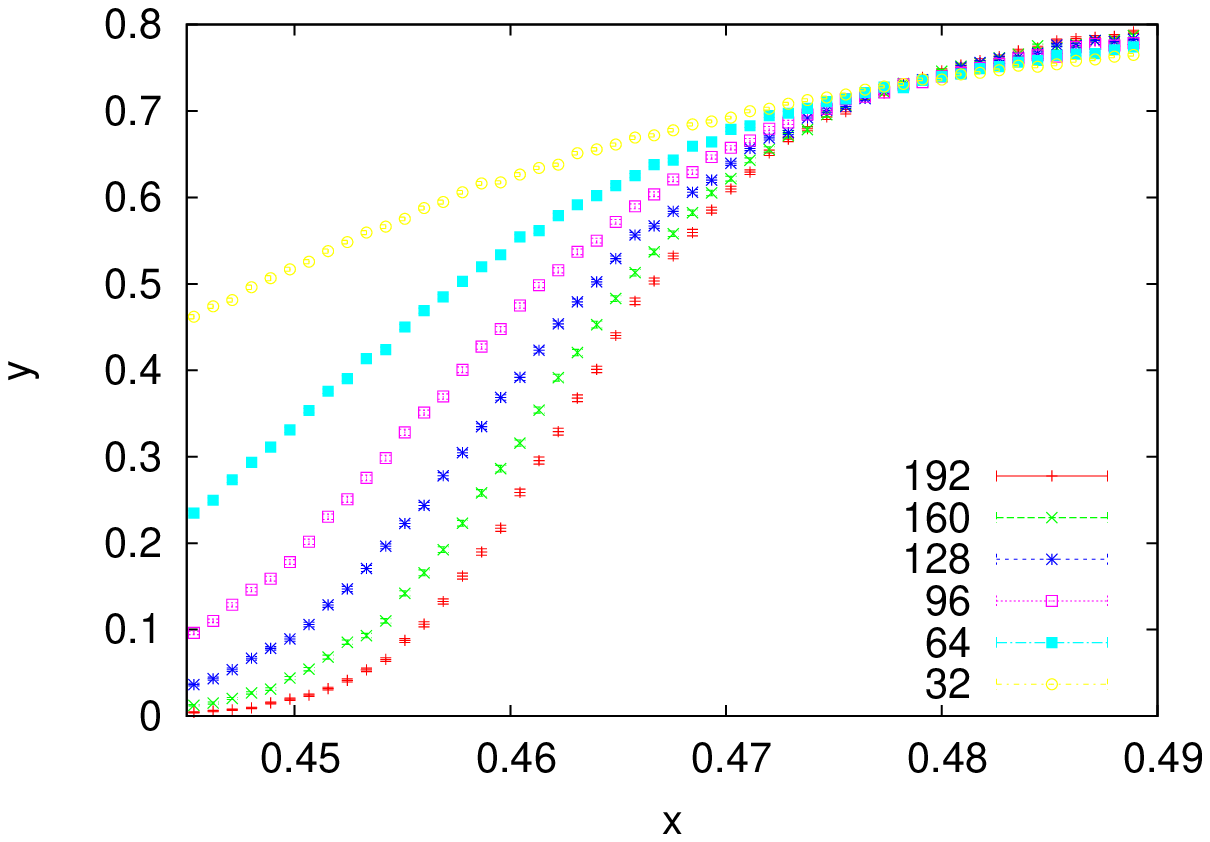}  \\
  \vspace{0.4cm}
\psfrag{x}[t][t][.7][0]{$(K'-K_\mathrm{c}')/L^{1/\nu}$}
  \includegraphics[width=.29\textwidth]{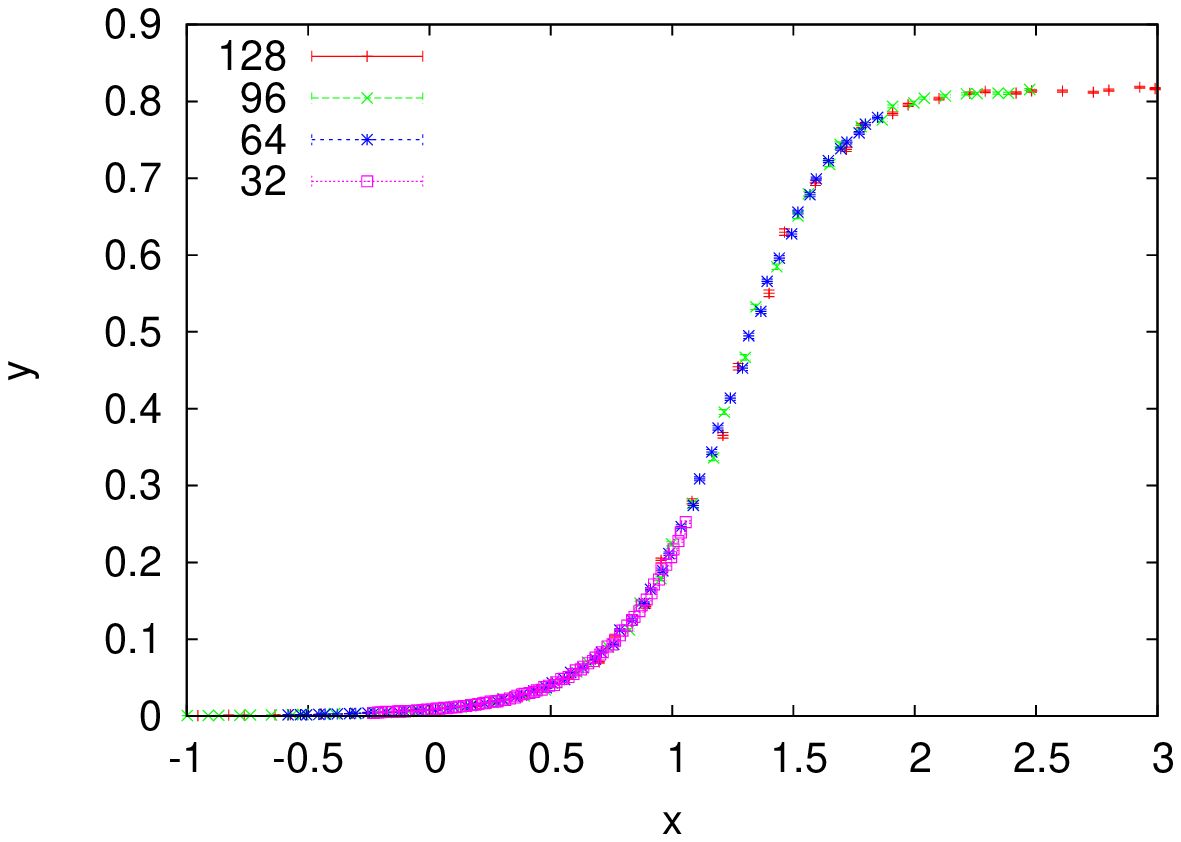} \hspace{.05\textwidth}
  \includegraphics[width=.29\textwidth]{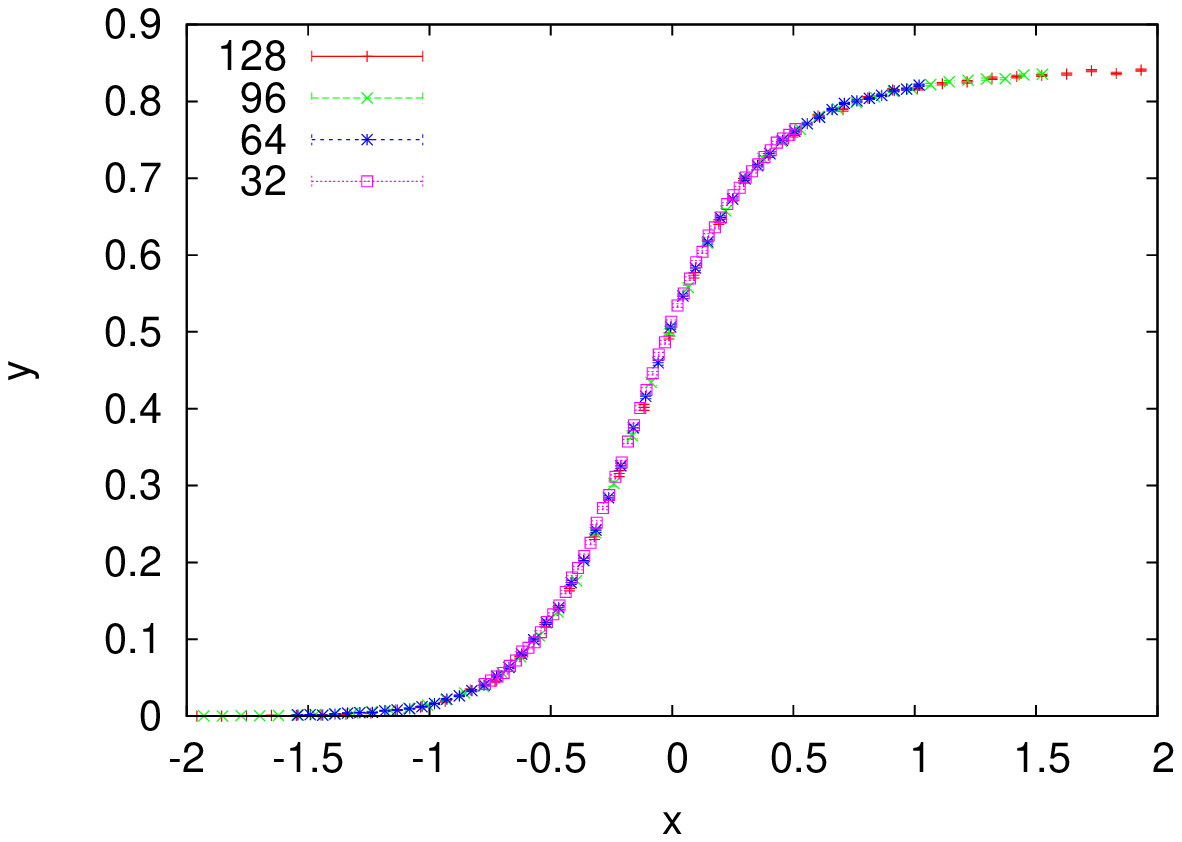}  \hspace{.05 \textwidth}
  \includegraphics[width=.29\textwidth]{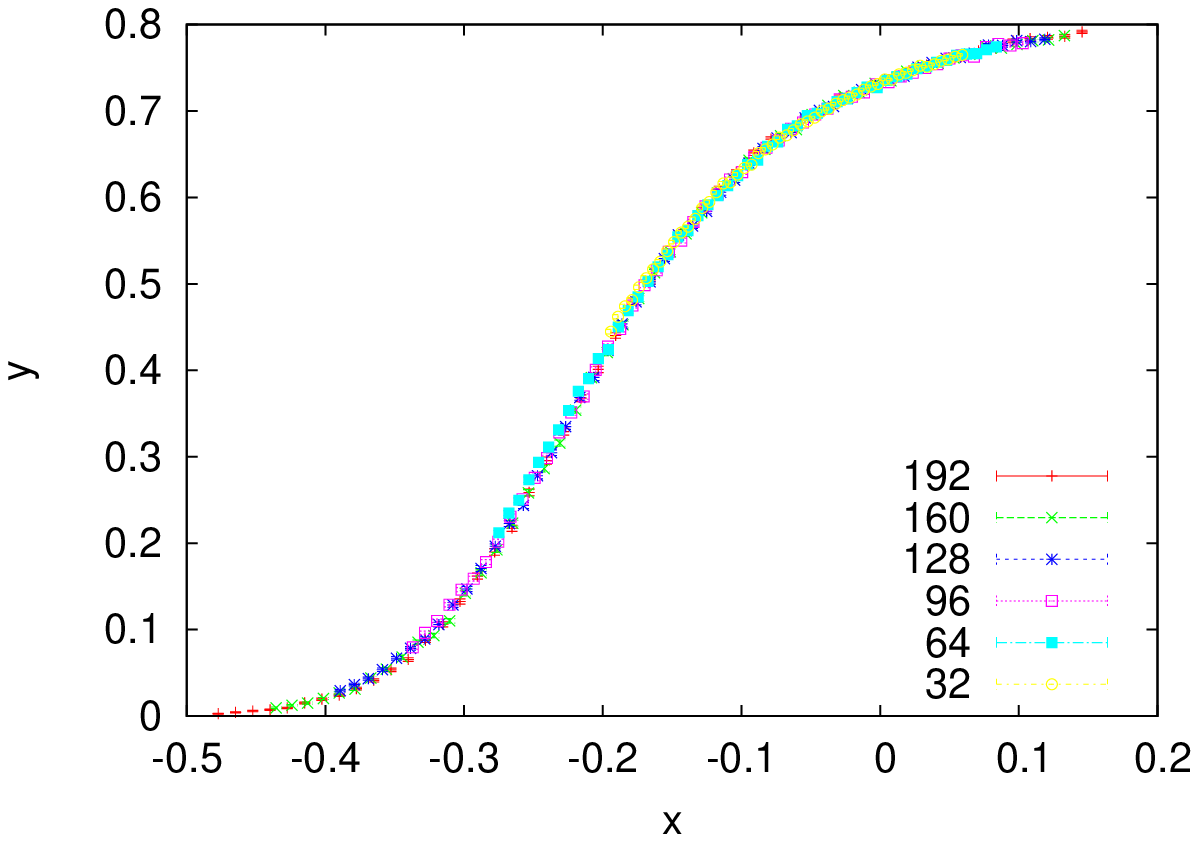} 
\end{center}
\caption{\emph{Top}: Probablity $\Pi_L$ that one or more loops wrap the
  lattice as a function of  $K'$ for various lattices of
  linear size $L$ and for $N=0.01,1$, and 1.8 (from left to right).
  \emph{Bottom}: Same data as above, but now plotted as a function of
  $(K'-K_\mathrm{c}')/L^{1/\nu}$ with the estimated values for
$K_\mathrm{c}'$ and the exactly known values for $\nu$.}
\label{fig:pi}
\end{figure*}

A backbite is
computationally the most costly update.  Because the
worm breaks into two parts, a new satellite list must be created to
accommodate either the detached loop or the remaining, shortened worm.
While the cost of cutting a segment out of the old satellite list and
pasting it into the new list is independent of the length of that
segment, the bonds belonging to the new list must be relabeled to
indicate that they are now part of a new list.  Since the computational
cost of relabeling increases linearly with the number of bonds, it is
expedient to move the shortest segment into the new list.  To identify
this part, we simultaneously iterate through the list, starting at each
of the two legs in front of the head of the worm until either the head
or the tail is reached.  After
inserting the shortest segment into the new list and relabeling the
bonds involved, we reconnect the loose ends at fixed cost, using the
unique properties of a satellite list.  

Using the algorithm outlined above, we determine the line of critical
points of the O($N$) loop model on the square lattice for $0 < N < 2$.
To locate the critical point for a given $N$, we measure the probability
$\Pi_L(K)$ that one or more loops wrap the lattice as a function of the
bond fugacity $K$ on systems of different linear size $L$.  This
percolation observable, shown in the top row of Fig.~\ref{fig:pi} for
$N=0.01, 1, 1.8$, has no scaling dimension so that, when plotted as a
function of $K$ or of the rescaled bond fugacity $K'\equiv K/N$,
the curves obtained for different $L$ should cross at exactly one point.
Being $L$-independent, this crossing point marks the value of $K$ for
which wrapping loops first appear on the \emph{infinite} lattice.  This
percolation threshold of HT loops coincides with the onset of long-range
spin-spin correlations  $K_\mathrm{c}$.  Figure~\ref{fig:pi} shows
that the crossing point can be accurately identified for the Ising model
($N=1$), but this task becomes increasingly more difficult as $N$
approaches either 0 or 2,  where the
crossing points move towards the plateaus at the beginning and at the
end of the curves, respectively.

To determine $K_\mathrm{c}$ more precisely, we use a second property of
$\Pi_L$, namely, that it is a function only of the ratio of the two
relevant length scales in the problem: the linear lattice size $L$ and
the correlation length $\xi$.  The latter diverges as $\xi \sim |K -
K_\mathrm{c}|^{-\nu}$, for $K\to K_\mathrm{c}$, where $\nu$ is the
correlation length exponent.  When replotted as a function of
$(K-K_\mathrm{c}) L^{1/\nu}$, the data should consequently collapse
onto a single curve.  Since $\nu$ is exactly known
for all $-2 \leq N \leq 2$ \cite{Nienhuis}, we estimate the location of
the threshold by keeping $\nu$ fixed and varying $K_\mathrm{c}$ until
the best data collapse is achieved. As an illustration, we plot in the
bottom row of Fig.~\ref{fig:pi} the best collapse achieved for the data
shown in the top row of the figure.  The quality of collapse is excellent 
for all values of $N$ we investigated in the range $0 < N
\lesssim 1.8$.  Table~\ref{table:Kc} 
summarizes our estimates obtained in this way.  The error bars represent the range in which the
quality of collapse, assessed by visual inspection, remains more or less
the same.  This simple method turns out to be very sensitive and yields
surprisingly precise results \cite{Winter}.  For $N \gtrsim 1.8$, the
quality diminishes, with the limiting case, $N \to 2$, where $1/\nu \to
0$, showing no collapse at all.

\begin{figure}
\psfrag{x}[t][t][.7][0]{$1/K$}
\psfrag{y}[t][t][.7][0]{$\langle w^2 \rangle$}
\begin{center}
\includegraphics[width=.4\textwidth]{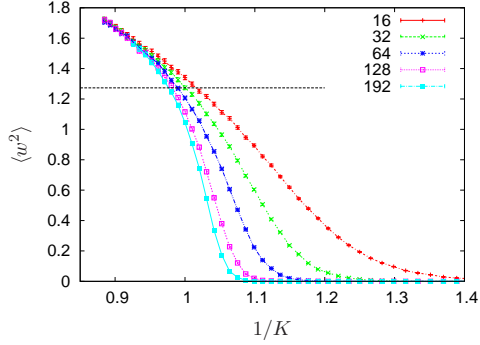}
\end{center}
\caption{The (scaled) helicity modulus $\langle w^2 \rangle$ as a
  function of $1/K$ for various $L$.  The horizontal line
  denotes the value $\langle w^2 \rangle = 4/\pi$. }
\label{fig:jump}
\end{figure}

\begin{table}
    \begin{tabular}{|l|l|l|l|l|l|}
        \hline
 $N$ & $K_\mathrm{c}'$ & $N$ & $K_\mathrm{c}'$ &  $N$ & $K_\mathrm{c}'$ \\ \hline 
0.01 &        0.37930(5)  &   0.7  &        0.4014(2)   & 1.4  &        0.4374(2)   \\ 
0.1  &        0.38178(3)  &   0.8  &        0.4053(1)   & 1.5  &        0.4450(2)   \\ 
0.2  &        0.38464(4)  &   0.9  &        0.4096(2)   & 1.6  &        0.4539(2)   \\ 
0.3  &        0.3877(4)   &   1.0  &        0.4141(1)   & 1.7  &        0.4647(2)   \\ 
0.4  &        0.3908(3)   &   1.1  &        0.4192(1)   & 1.8  &        0.4785(2)   \\ 
0.5  &        0.3941(1)   &   1.2  &        0.4246(2)   & 2.0  &        0.520(4)    \\
0.6  &        0.3977(2)   &   1.3  &        0.4307(1)   &      &                    \\   
   \hline
    \end{tabular}
\caption{Estimates of 
$K_\mathrm{c}'$ for various values of $N$.}
\label{table:Kc}
\end{table}
To investigate this issue, we consider in detail the limiting case
$N=2$, corresponding to the XY model. Besides logarithmic corrections to scaling, the XY
critical point is special also because the correlation length diverges
exponentially instead of powerlike as happens for $N<2$
\cite{Kosterlitz}.  These two complicating factors make it notoriously
difficult to determine $K_c$ for $N=2$ and  usually
require Monte Carlo simulations on lattices of sizes much larger than
those used for $N<2$ to achieve comparable precisions.  One way to
circumvent this problem \cite{Minnhagen1987} is by analyzing the
size-dependence of the  helicity modulus
$\Upsilon$, which exhibits a universal jump at the critical point that
is unique to this the BKT phase transition
\cite{Nelson1977}.  In the HT representation, the helicity modulus
 is determined by the average squared winding
number of loop configurations, $\langle w^2 \rangle$, where $w$ is the
number of times the loops wrap the lattice (in any of the two
directions) for a given loop configuration \cite{Ceperley1986}.  As
shown in Fig.~\ref{fig:jump},  this observable has the
 expected behavior,
namely, it is finite in the low-temperature phase and rapidly falls to
zero when $1/K$ increases.

\begin{figure}
\psfrag{x}[t][t][.7][0]{$L$}
\psfrag{y}[t][t][.7][0]{$-1/x$}
\begin{center}
\includegraphics[width=.4\textwidth]{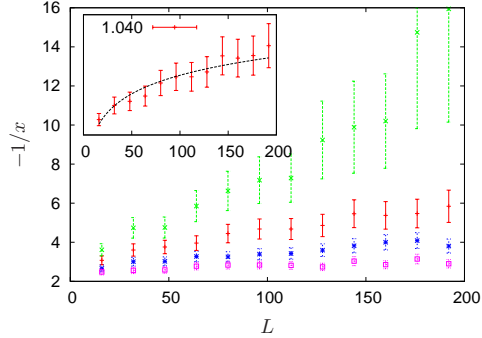}
\end{center}
\caption{The negative of the inverse of the renormalization group variable $x$,
  defined in Eq.~(\ref{x}), as a function of lattice size $L$ for $K =
  1.030, 1.040, 1.050, 1.060$ (from top to bottom).  The curve in the
  inset is the best fit to the data at $K=1.040$.}
\label{fig:fit}
\end{figure}

The quantity $\langle w^2 \rangle$ features in the Kosterlitz
renormalization group equations \cite{Kosterlitz} through the
combination
\begin{equation} 
\label{x}
x \equiv 2 - \frac{\pi}{2} \langle w^2 \rangle .
\end{equation}   
For the infinite system, one has $\lim_{K \to K_\mathrm{c}^+} \langle
w^2 \rangle ={4}/{\pi}$, so that $x$ vanishes at the critical point in
the thermodynamic limit.  On a finite lattice, $x$ remains finite at the
critical point and its size dependence is known \cite{Minnhagen1987} to
be given by
\begin{equation} 
\label{sizedep}
\lim_{K \to K_\mathrm{c}^+} x = - \frac{1}{\ln(L/L_0)},
\end{equation} 
where $L_0$ is a characteristic length of the order of the lattice spacing.
Relation (\ref{sizedep}) can be used to estimate the critical point of
the infinite lattice as follows \cite{Minnhagen1987}.  First one
measures $x$ on lattices of
different sizes for different values of $K$ and then  fits the size-dependence
 (\ref{sizedep}) to the data
  for each $K$, using $L_0$ as the
only free parameter, see Fig.~\ref{fig:fit}.  The value of $K$ that
produces the best fit is taken as estimate of the critical point of
the infinite lattice.  We in this way arrive at the estimate
$K_\mathrm{c} = 1.040(7)$ and $L_0 = 0.88(5)$, for which
$\chi^2/\textsc{dof} = 0.25$, see inset of Fig.~\ref{fig:fit}.

In conclusion, taking full account of crossing, we have simulated the
$O(N)$ loop model on the square lattice with the worm algorithm.  By
representing the worm and the loops it generates by satellite lists, we
have succeeded to monitor the number of loops present on the lattice, a
non-local quantity, throughout the simulations in a very efficient way.
This enabled us to determine the critical point (and other properties)
of the $O(N)$ model on the square lattice for arbitrary $0<N\le 2$. We
 presently investige the behavior of the model
in the low-temperature phase for $0<N\le 2$ and that for $N>2$, where no
transition associated with the appearance of loops wrapping the infinite
lattice is expected.  The algorithm presented here is by no means
  restricted to the square lattice, and can be used to simulate the
  O($N$) model for arbitrary $N>0$ on any (higher-dimensional) lattice.

\begin{acknowledgments}
The work of A.M.J.S. was financialy supported by CAPES, Brazil through a
visiting professor scholarship.  That author would like to thank
Wolfhard Janke for helpful discussions. A.M.P.S and G.L.V. acknowledge
financial support from CNPq and FACEPE (Brazilian agencies).
\end{acknowledgments}

\end{document}